\documentclass[showpacs,preprintnumbers,superscriptaddress]{revtex4}
\usepackage{CJK}
\usepackage{amsmath,amssymb,graphicx,bm}
\begin{document}

\title{Effects of Lorentz breaking on the accretion onto a Schwarzschild-like black hole}
\author{Rongjia Yang \footnote{Corresponding author}}
\email{yangrongjia@tsinghua.org.cn}
\affiliation{College of Physical Science and Technology, Hebei University, Baoding 071002, China}
\affiliation{Hebei Key Lab of Optic-Electronic Information and Materials, Hebei University, Baoding 071002, China}
\affiliation{National-Local Joint Engineering Laboratory of New Energy Photoelectric Devices, Hebei University, Baoding 071002, China}
\author{He Gao}
\affiliation{College of Physical Science and Technology, Hebei University, Baoding 071002, China}
\author{Yaoguang Zheng}
\affiliation{College of Physical Science and Technology, Hebei University, Baoding 071002, China}
\author{Qin Wu}
\affiliation{School of information engineering, Guangdong Medical University, Dongguan 523808, China}

\begin{abstract}
We formulate and solve the problem of spherically symmetric, steady state, adiabatic accretion onto
a Schwarzschild-like black hole obtained recently. We derive the general analytic expressions for the critical points, the critical velocity, the
critical speed of sound, and subsequently the mass accretion rate. The case for polytropic gas is discussed in detail. We find the parameter characterizing the breaking of Lorentz symmetry will slow down the mass accretion rate, while has no effect on the gas compression and the temperature profile below the critical radius and at the event horizon.
\end{abstract}


\maketitle

\section{Introduction}

Accretion of matter onto black hole is the most likely scenario to explain
the high energy output from active galactic nuclei and quasars, which is an important phenomenon of long-standing interest to astrophysicists.
Pressure-free gas dragged onto a massive central object was first considered in \cite{hoyle1939effect,Bondi:1944jm}, which was generalized to
the case of the spherical accretion of adiabatic fluids onto astrophysical objects \cite{Bondi:1952ni}. In the framework of general relativity
the steady-state spherical symmetric flow of matter into or out of a condensed object was examined by Michel \cite{michel1972accretion}. Since then accretion has been an extensively
studied topic in the literature, including accretion onto a Schwarzschild black hole \cite{Babichev:2004yx,Paik:2017wcy}, onto a Reissner-Nordstrom black hole \citep{Babichev:2008jb, michel1972accretion, Jamil:2008bc, Rodrigues:2016uor}, onto a Kerr-Newman black hole \citep{Babichev:2008dy, JimenezMadrid:2005rk,Bhadra:2011me}, onto a Kiselev black hole \cite{Yang:2016sjy}, and onto a black hole in a string cloud background \citep{Ganguly:2014cqa}, onto cosmological black holes or onto Schwarzschild-(anti-)de Sitter spacetimes \citep{Gao:2008jv, Mach:2013fsa, Mach:2013gia, Karkowski:2012vt}. Quantum gravity effects of accretion onto a Schwarzschild black hole were considered in the context of asymptotically safe scenario \cite{Yang:2015sfa}. An exact solution was obtained for dust shells collapsing towards a black hole \cite{Liu:2009ts}, which was generalized to the case when the tangential pressure is taken into account \cite{Zhao:2018ani}. Analytic solutions for accretion of a gaseous medium with a adiabatic equation of state ($P=\rho$) were obtained for a moving Schwarzschild black hole, for a moving Kerr back hole \cite{petrich1988accretion}, and for a moving Reissner-Nordstr\"{o}m black hole \cite{Jiao:2016uiv}.


An interesting study of higher dimensional accretion onto TeV black holes was discussed in the Newtonian limit \cite{Giddings:2008gr}. The accretion of phantom matter onto 5-dimensional charged black holes was studied in \cite{Sharif:2011ih}. The accretion of phantom energy onto 5-dimensional extreme Einstein-Maxwell-Gauss-Bonnet black hole was investigated in \cite{Jamil:2011sx}. Accretions were analyzed in higher-dimensional Schwarzschild black hole \cite{John:2013bqa} and in higher-dimensional Reissner-Nordstr\"{o}m black hole \cite{Yang:2018cim, Sharif:2016pqy}. Accretions of Dark Matter and Dark Energy onto ($n+2$)-dimensional Schwarzschild Black Hole and Morris-Thorne wormhole was considered in \cite{Debnath:2015yva}. We will determine analytically the critical points, the critical fluid velocity and the critical sound speed, and subsequently the mass accretion rate. More interesting, we will investigate whether the parameter characterizing the breaking of Lorentz symmetry has effects on the accretion.


The organization of this paper is as follows. In section II, we will present the fundamental equations for the accretion of matter
onto a Schwarzschild-like black hole. In section III, we will consider the critical points and the conditions
the critical points must satisfy. In section IV, we will discuss the case of polytropic gas in detail as an application.
Finally, We will briefly summarize and discuss our results in section V.

\section{Basic equations for accretion}
A static and spherically symmetric exact solution of the Einstein equations, called Schwarzschild-like black hole, in the presence of a spontaneous breaking of Lorentz symmetry because of the nonzero vacuum expectation value of the bumblebee field was obtained in \cite{Casana:2017jkc}. The geometry of Schwarzschild-like black hole is given by the following line element
\begin{eqnarray}
\label{met}
ds^{2}=-\left(1-\frac{2M}{r}\right)dt^{2}+(1+l)\left(1-\frac{2M}{r}\right)^{-1}dr^{2}+r^{2}d\theta^{2}+r^{2}sin^{2}\theta d\phi ^{2}
\end{eqnarray}
where $l$ is a constant characterizing the breaking of Lorentz symmetry and $M$ is the mass of the black hole. We chose the unit system as $G=c=1$ throughout the paper. It is easy to see that the horizon of black hole is at $r_{\rm {H}}=2M$.

We consider a static radial matter flow onto the black hole without back-reaction. The flow is an ideal fluid which is approximated by the following energy momentum tensor
\begin{eqnarray}
T^{\alpha\beta}&=&(\rho+p)u^{\alpha}u^{\beta}+pg^{\alpha\beta},
\end{eqnarray}
where $\rho$ is the proper energy density and $p$ is the proper pressure of the fluid. $u^{\alpha}=dx^{\alpha}/ds$ is the 4-velocity of the fluid, which obeys the normalization condition $u^{\alpha}u_{\alpha}=-1$.


We define the radial component of the four-dimensional velocity as $\upsilon(r)=u'=\frac{dr}{ds}$. Since the velocity component is zero for $\alpha>1$, we get from the normalization condition
\begin{eqnarray}
(u^{0})^{2}=\frac{(1+l)\upsilon^{2}+1-\frac{2M}{r}}{\left(1-\frac{2M}{r}\right)^{2}}.
\end{eqnarray}
If ignoring the self-gravity of the flow, one can define baryon numbers density $n$ and numbers flow $J^{\alpha}=nu^{\alpha}$ in the flow's local inertial frame. Assuming that no particles are generated or disappeared, then the number of particles is conserved, we have
\begin{eqnarray}
\label{part}
\nabla_{\alpha}J^{\alpha}=\nabla_{\alpha}(nu^{\alpha})=0,
\end{eqnarray}
where $\nabla_{\alpha}$ represents covariant derivative with respect to the coordinate $x^{\alpha}$.  For the metric (\ref{met}),
the equation (\ref{part}) can be rewritten as
\begin{eqnarray}
\label{nine}
\frac{1}{r^{2}}\frac{d}{dr}\left( r^{2}n\upsilon\right)=0,
\end{eqnarray}
which for a perfect fluid gives the integration as
\begin{eqnarray}
\label{9a}
r^{2}n\upsilon=C_1,
\end{eqnarray}
where $C_1$ is an integration constant. Integrating the equation (\ref{nine}) over four-dimensional volume and multiplying with the mass
of each particle, $m$, we have
\begin{eqnarray}
\label{accret}
\dot{M}=4\pi r^{2}mn\upsilon,
\end{eqnarray}
where $\dot{M}$ is an integration constant which has the dimension of mass per unit time. This is the Bondi mass accretion rate actually. Ignoring the back-reaction,  the $\beta=0$ component of the energy-momentum conservation, $\nabla_{\alpha}T^{\alpha}_{\beta}=0$, gives
for spherical symmetry and steady-state flow
\begin{eqnarray}
\label{ten}
\frac{1}{r^{2}}\frac{d}{dr}\left[r^{2}(\rho+p)\upsilon\sqrt{1-\frac{2M}{r}+(1+l)\upsilon^{2}}\right]=0,
\end{eqnarray}
which can be integrated as
\begin{eqnarray}
\label{10a}
r^{2}(\rho+p)\upsilon\sqrt{1-\frac{2M}{r}+(1+l)\upsilon^{2}}=C_2,
\end{eqnarray}
where $C_2$ is an integration constant. Dividing Eqs. (\ref{10a}) and (\ref{9a}) and then squaring, we derive
\begin{eqnarray}
\label{elev}
\left(\frac{\rho+p}{n}\right)^{2}\left[1-\frac{2M}{r}+(1+l)\upsilon^{2}\right]=\left(\frac{\rho_{\infty}+p_{\infty}}{n_{\infty}}\right)^{2}.
\end{eqnarray}
The $\beta=1$ component of the energy-momentum conservation, $\nabla_{\alpha}T^{\alpha}_{\beta}=0$, yields
\begin{eqnarray}
\label{tew}
(1+l)\upsilon\frac{d\upsilon}{dr}=-\frac{dp}{dr}\frac{1-\frac{2M}{r}+(1+l)\upsilon^{2}}{\rho+p}-\frac{M}{r^{2}}.
\end{eqnarray}
The equations (\ref{accret})and (\ref{elev}) are the basic conservation equations for the material flow onto a four-dimensional schwarzschild-like black hole (\ref{met}) when the back-reaction of matter is ignored.

\section{Conditions for critical accretion}
In this section, we consider the conditions for critical accretion. In the local inertial rest frame of the fluid, the conservation of mass-energy for adiabatic flow without entropy generation is governed by
\begin{eqnarray}
\label{mass}
0=Tds=d\left(\frac{\rho}{n}\right)+pd\left(\frac{1}{n}\right),
\end{eqnarray}
from which we can get the following relationship
\begin{eqnarray}
\frac{d\rho}{dn}=\frac{\rho+p}{n}.
\end{eqnarray}
Using this equation, we define the speed of adiabatic sound as
\begin{eqnarray}
\label{six}
a^{2}\equiv\frac{dp}{d\rho}=\frac{n}{\rho+p}\frac{dp}{dn}.
\end{eqnarray}
Differentiating Eqs. (\ref{9a}) and (\ref{tew}) with respect to $r$, which gives, respectively
\begin{eqnarray}
\frac{1}{\upsilon}\upsilon '+\frac{1}{n}n'=-\frac{2}{r},
\end{eqnarray}
\begin{eqnarray}
(1+l)\upsilon\upsilon '+\left[1-\frac{2M}{r}+(1+l)\upsilon^{2}\right]a^{2}\frac{n'}{n}=-\frac{M}{r^{2}},
\end{eqnarray}
where the prime represents a derivative with respect to $r$. From these equations we get the following system
\begin{eqnarray}
\label{sixt}
\upsilon'=\frac{N_{1}}{N},
\end{eqnarray}
\begin{eqnarray}
\label{seven}
n'=-\frac{N_{2}}{N},
\end{eqnarray}
where
\begin{eqnarray}
N_{1}=\frac{1}{n}\left\{\left[1-\frac{2M}{r}+(1+l)\upsilon^{2}\right]\frac{2a^{2}}{r}-\frac{M}{r^{2}}\right\},
\end{eqnarray}
\begin{eqnarray}
N_{2}=\frac{1}{\upsilon}\left[\frac{2(1+l)\upsilon^{2}}{r}-\frac{M}{r^{2}}\right],
\end{eqnarray}
\begin{eqnarray}
\label{nin}
N=\frac{(1+l)\upsilon^{2}-\left[1-\frac{2M}{r}+(1+l)\upsilon^{2}\right]a^{2}}{n\upsilon}.
\end{eqnarray}
For large values of $r$,  we require the flow to be subsonic, that is $\upsilon<a$. Because the speed of sound is always less than the speed of light, $a<1$, then we have $\upsilon^2\ll 1$. The equation (\ref{nin}) can be approximated as
\begin{eqnarray}
N\approx\frac{(1+l)\upsilon^{2}-a^{2}}{n\upsilon}
\end{eqnarray}
and since $0<l\ll 1$ and $v<0$, we have $N>0$ as $r\rightarrow\infty$. In the event horizon, $r_{\rm{H}}=2M$, we can get
\begin{eqnarray}
N=\frac{(1+l)\upsilon^{2}(1-a^{2})}{n\upsilon}
\end{eqnarray}
Under the constraint of causality, $a^{2}<1$, we have $N<0$. Therefore, there must exists critical points $r_{\rm{c}}$ where $r_{\rm{H}}<r_{\rm{c}}<\infty$, at which $N=0$. The flow must pass the critical points outside the horizon, to avoid the appearance of discontinuity in the flow, we must require $N=N_{1}=N_{2}=0$ at $r=r_{\rm {c}}$, namely
\begin{eqnarray}
\label{hava}
N_{1}=\frac{1}{n_{\rm {c}}}\left\{\left[1-\frac{2M}{r_{\rm {c}}}+(1+l)\upsilon_{\rm {c}}^{2}\right]\frac{2a_{\rm {c}}^{2}}{r_{\rm {c}}}-\frac{M}{r_{\rm {c}}^{2}}\right\}=0,
\end{eqnarray}
\begin{eqnarray}
\label{havb}
N_{2}=\frac{1}{\upsilon_{\rm {c}}}\left[\frac{2(1+l)\upsilon_{\rm {c}}^{2}}{r_{\rm {c}}}-\frac{M}{r_{\rm {c}}^{2}}\right]=0,
\end{eqnarray}
\begin{eqnarray}
\label{havc}
N=\frac{(1+l)\upsilon_{\rm {c}}^{2}-\left[1-\frac{2M}{r_{\rm {c}}}+(1+l)\upsilon_{\rm {c}}^{2}\right]a_{\rm {c}}^{2}}{n_{\rm {c}}\upsilon_{\rm {c}}}=0,
\end{eqnarray}
where $a_{\rm {c}}\equiv a(r_{\rm {c}})$, $\upsilon_{\rm {c}}\equiv\upsilon(r_{\rm {c}})$, etc. From equation (\ref{hava}), (\ref{havb}) and (\ref{havc}), we obtain the radial velocity, the speed of sound at the critical points, and the critical radius, respectively
\begin{eqnarray}
\label{radial}
\label{vc}
\upsilon_{\rm {c}}^{2}=\frac{(2r_{\rm {c}}-3M)a^{2}_{\rm {c}}}{2(1+l)r_{\rm {c}}}=\frac{M}{2(1+l)r_{\rm {c}}},
\end{eqnarray}
\begin{eqnarray}
\label{sound}
a_{\rm {c}}^{2}=\frac{M}{2r_{\rm {c}}-3M},
\end{eqnarray}
\begin{eqnarray}
\label{rc}
r_{\rm {c}}=\frac{(3a^{2}_{\rm {c}}+1)M}{2a^{2}_{\rm {c}}}.
\end{eqnarray}
For $\upsilon_{\rm {c}}^{2}\geq 0$ and $a_{\rm {c}}^{2}\geq 0$, the equations (\ref{sixt}) and (\ref{seven}) exist acceptable solutions, therefore we have
\begin{eqnarray}
\left\{
\begin{aligned}
\upsilon_{\rm {c}}^{2}&=\frac{M}{2(1+l)r_{\rm {c}}}\geq0, \\
a_{\rm {c}}^{2}&=\frac{M}{2r_{\rm {c}}-3M}\geq0,
\end{aligned}
\right.
\end{eqnarray}
from which we can obtain the conditions satisfied by the critical points
\begin{eqnarray}
r_{\rm {c}}\geq\frac{3}{2}M.
\end{eqnarray}
This equation can make Eqs.(\ref{radial}) and (\ref{sound}) exist physical solutions. However, taking into account the restrictions of causality, $a^{2}_{\rm{c}}<1$, we have $r_{\rm{c}}>r_{\rm{H}}$ from (\ref{rc}), meaning that the critical points are located at outside of the event horizon.

\section{The polytropic solution}
In this section, we will analysis the accretion rate for polytrope gas and compute the gas compression and the adiabatic temperature distribution at the outer horizon.

\subsection{Accretion for polytrope gas}
For an example to calculate explicitly $\dot{M}$, we consider the polytrope introduced in \cite{Bondi:1944jm, Bondi:1952ni} with the equation of
state
\begin{eqnarray}
\label{poly}
p=Kn^{\gamma},
\end{eqnarray}
where $K$ is a constant and $\gamma$ is adiabatic index satisfing $1<\gamma<\frac{5}{3}$. Substitute this expression into the equation for conservation of mass-energy and integrating, we can obtain
\begin{eqnarray}
\label{get}
\rho=\frac{K}{\gamma-1}n^\gamma+mn,
\end{eqnarray}
where $mn$ is the rest energy density with $m$ an integration constant. With the definition of the speed of sound (\ref{six}), we can rewritten the Bernoulli equation (\ref{elev}) as
\begin{eqnarray}
\label{lab}
\left(1+\frac{a^{2}}{\gamma-1-a^{2}}\right)^{2}\left[1-\frac{2M}{r}+(1+l)\upsilon^{2}\right]=\left(1+\frac{a_{\infty}^{2}}{\gamma-1-a_{\infty}^{2}}\right)^{2}.
\end{eqnarray}
According to the critical radial velocity (\ref{radial}) and the critical speed of sound (\ref{sound}), the equation (\ref{lab}) must satisfy the condition at the critical point $r_{\rm{c}}$,
\begin{eqnarray}
\label{34}
(1+3a_{\rm{c}}^{2})\left(1-\frac{a_{\rm{c}}^{2}}{\gamma-1}\right)^{2}=\left(1-\frac{a_{\infty}^{2}}{\gamma-1}\right)^{2}.
\end{eqnarray}
Since the baryons are still non-relativistic for $r_{\rm{c}}<r<\infty$, one can expect that $a^{2}<a_{c}^{2}\ll1$. Expanding (\ref{34}) to the first order
in $a_{\rm{c}}^{2}$ and $a_{\infty}^{2}$, yields
\begin{eqnarray}
a_{\rm{c}}^{2}\approx\frac{2}{5-3\gamma}a_{\infty}^{2},
\end{eqnarray}
with this equation, we can get the expression of the critical points $r_{\rm{c}}$ in terms of the boundary condition $a_{\infty}$ and the black-hole mass $M$ as
\begin{eqnarray}
\label{rc1}
r_{\rm{c}}\approx\frac{(5-3\gamma)M}{4a_{\infty}^{2}}.
\end{eqnarray}
From the equations (\ref{six}), (\ref{poly}), and (\ref{get}), we have
\begin{eqnarray}
\gamma K n^{\gamma-1}=\frac{ma^{2}}{1-\frac{a^{2}}{\gamma-1}}.
\end{eqnarray}
Since $a^{2}\ll 1$, we have $n\sim a^{\frac{2}{\gamma-1}}$ and
\begin{eqnarray}
\label{nc}
\frac{n_{\rm{c}}}{n_{\infty}}=\left(\frac{a_{\rm{c}}}{a_{\infty}}\right)^{\frac{2}{\gamma-1}}.
\end{eqnarray}
Since $\dot{M}$ is independent of $r$ implying from the equation (\ref{accret}), it must also hold at $r=r_{\rm{c}}$, with which we can determine the accretion rate
\begin{eqnarray}
\label{m1}
\dot{M}&=&4\pi r^{2}m n\upsilon=4\pi r_{\rm{c}}^{2}m n_{\rm{c}}\upsilon_{\rm{c}}.
\end{eqnarray}
Substituting the equations (\ref{vc}), (\ref{rc1}), and (\ref{nc}) into the equation (\ref{m1}), we derive
\begin{eqnarray}
\label{acc1}
\dot{M}=4\pi \left(\frac{M}{a_{\infty}^{2}}\right)^{2}m n_{\infty}a_{\infty}\left(\frac{1}{1+l}\right)^{\frac{1}{2}}\left(\frac{1}{2}\right)^{\frac{\gamma+1}{2(\gamma-1)}}\left(\frac{5-3\gamma}{4}\right)^{\frac{3\gamma-5}{2(\gamma-1)}}.
\end{eqnarray}
For $l=0$, the equation (\ref{acc1}) reduces to the results in Schwarzschild black hole case. The parameter $l$ can slows down the mass accretion rate. For example, comparing with the total mass accretion rate, the decrease of the mass accretion rate resulting from the Lorentz breaking parameter $l$ is about $10^{-7}$ orders, for the parameter $l$ bounded as $l<10^{-13}$ \cite{Casana:2017jkc}. This may be a valuable feature to incorporate in astrophysical applications.

\subsection{Asymptotic behavior at the event horizon}
In the previous sections we obtained the results are valid at large distances from the black hole near the critical radius $r_{\rm{c}}\gg r_{H}$. Now, we analysis the flow characteristics for $r_{\rm{H}}<r\ll r_{c}$ and at the even horizon $r=r_{\rm{H}}$.

For $r_{\rm{H}}<r\ll r_{c}$, we obtain the fluid velocity from Eq. (\ref{lab}), it can be approximated as
\begin{eqnarray}
\label{last}
\upsilon^{2}\approx\frac{2M}{(1+l)r}.
\end{eqnarray}
Usin the equations (\ref{accret}), (\ref{acc1}),  and (\ref{last}), we can estimate the gas compression on these scales
\begin{eqnarray}
\label{nn}
\frac{n(r)}{n_{\infty}}=\left(\frac{M}{a^{2}_{\infty}r}\right)^{\frac{3}{2}}
\left(\frac{1}{2}\right)^{\frac{2\gamma}{2(\gamma-1)}}\left(\frac{5-3\gamma}{4}\right)^{\frac{3\gamma-5}{2(\gamma-1)}}.
\end{eqnarray}
Assuming a Maxwell-Boltzmann gas with $p=n\kappa_{\rm{B}}T$, we get the adiabatic temperature profile by using (\ref{poly}) and (\ref{nn})
\begin{eqnarray}
\label{tr}
\frac{T(r)}{T_{\infty}}=\left(\frac{M}{a^{2}_{\infty}r}\right)^{\frac{3(\gamma-1)}{2}}
\left(\frac{1}{2}\right)^{\gamma}\left(\frac{5-3\gamma}{4}\right)^{\frac{3\gamma-5}{2}}.
\end{eqnarray}
Near the outer horizon, the flow is supersonic, and the fluid velocity is still well approximated by (\ref{last}). At the horizon $r=r_{\rm{H}}=2M$, the flow velocity is approximately equal to $\upsilon^{2}\approx\frac{1}{1+l}$. We can derive the gas compression at the horizon by using the equations (\ref{accret}), (\ref{acc1}), and (\ref{last})
\begin{eqnarray}
\label{nh}
\frac{n_{\rm{H}}}{n_{\infty}}=\frac{1}{a^{3}_{\infty}}\left(\frac{1}{2}\right)^{\frac{5\gamma-3}{2(\gamma-1)}}\left(\frac{5-3\gamma}{4}\right)^{\frac{3\gamma-5}{2(\gamma-1)}}.
\end{eqnarray}
Using equations (\ref{poly}) and (\ref{nh}), we obtain the adiabatic temperature profile at the event horizon for a Maxwell-Boltzmann gas with $p=n\kappa_{\rm{B}}T$
\begin{eqnarray}
\label{th}
\frac{T_{\rm{H}}}{T_{\infty}}=\left(\frac{1}{a^{3}_{\infty}}\right)^{\gamma-1}
\left(\frac{1}{2}\right)^{\frac{5\gamma-3}{2}}\left(\frac{5-3\gamma}{4}\right)^{\frac{3\gamma-5}{2}}.
\end{eqnarray}
Equations (\ref{nn}), (\ref{tr}), (\ref{nh}), and (\ref{th}) show that the parameter $l$ have no effects on the gas compression and the adiabatic temperature profile, so these equations are the same in Schwarzschild black hole case.

\section{Conclusions and discussions}
In this paper we formulated and solved the problem of spherically symmetric, steady state, adiabatic accretion onto
a Schwarzschild-like black hole in four-dimensional spacetime by using four-dimensional general relativity. We
obtained the general analytic expressions for the critical points, the critical velocity of fluid, the
critical speed of sound, and subsequently the mass accretion rate. We determined the
physical conditions the critical points should satisfy. We obtained the explicit expressions for the gas compression and the temperature profile below the critical
radius and at the event horizon. We found the parameter characterizing the breaking of Lorentz symmetry will slows down the mass accretion rate, while has no effect on the gas compression and the temperature profile below the critical radius and at the event horizon. These results useful for
feature researches.

\begin{acknowledgments}
This study is supported in part by National Natural Science Foundation of China (Grant No. 11273010), Hebei Provincial Natural Science Foundation of China (Grant No. A2014201068), the Outstanding Youth Fund of Hebei University (No. 2012JQ02), and the Midwest universities comprehensive strength promotion project.
\end{acknowledgments}

\bibliographystyle{elsarticle-num}
\bibliography{ref}

\end{document}